\documentclass{lhep}       

\usepackage[OT1]{fontenc}%for font support
\usepackage[hyperfootnotes=false]{hyperref}
\usepackage{amsmath}
\usepackage{amssymb}
\usepackage{gensymb}
\usepackage{microtype}
\usepackage{lipsum}

\journal{LHEP}%Letters in High Energy Physics}
\vol{xx}
\jyear{2018}
\pages{xxx} 

\received{xx January 2018}
\published{xx March 2018}

\def\be{\begin{equation}}
\def\ee{\end{equation}}
\def\bea{\begin{eqnarray}}
\def\eea{\end{eqnarray}}

\renewcommand{\d}{\mathrm{d}}
\newcommand{\diff}[2]{\dfrac{\mathrm{d} #1}{\mathrm{d} #2}}
\newcommand{\inlinediff}[2]{\mathrm{d}{#1}/\mathrm{d}{#2}}
\newcommand{\pdiff}[2]{\dfrac{\partial #1}{\partial #2}}
\newcommand{\cs}{\langle \sigma v \rangle}

\newcommand{\RXC}{RXC~J0225.1-2928}

\begin{document}

% \title{Accurate modelling of the synchrotron radio emissions from WIMP dark matter}
\title{A radio-frequency search for WIMPs in \RXC{}}

\author{Michael Sarkis,\auno{1} Geoff Beck,\auno{1} and Natasha Lavis,\auno{1}}
\address{$^1$School of Physics, University of the Witwatersrand, Private Bag 3, WITS-2050, Johannesburg, South Africa}

% \author{}
% \address{}

\begin{abstract}
Recent studies focusing on the use of radio data in indirect dark matter detection have led to a set of highly competitive limits on the WIMP annihilation cross-section, especially in light of high-resolution data from instruments like ASKAP and MeerKAT. In this work we present an analysis of radio observations of the \RXC{} galaxy cluster, taken from the recent MeerKAT Galaxy Cluster Legacy Survey public data release. We adopt a robust morphological analysis of this source that allows us to derive a set of upper-limits on the annihilation cross-section, and in our most constraining scenario these results are comparable to the most stringent limits yet found in the literature. 
\end{abstract}

\maketitle


\section{Introduction}

Our current standard cosmological model, $\Lambda$CDM, requires that a significant portion of the universe's energy density be in the form of Dark Matter (DM). A large part of our search for this elusive substance has been through indirect detection methods -- those which try to detect astrophysical signals emanating from the products of DM particle interactions. Some notable recent studies~\cite{beckRadioFrequencySearchesDark2019,chanRadioConstraintsDark2021,regisEMUViewLarge2021,egorovUpdatedConstraintsWIMP2022} have renewed interest in the use of radio observations for DM searches, which relies on the emission of synchrotron radiation by charged annihilation products (usually electrons and positrons) travelling within the magnetised environments of various astronomical objects. 

The results of these recent radio analyses are a set of very stringent limits on the DM annihilation cross-section; the authors of~\cite{regisEMUViewLarge2021} for example quote the strongest limits yet found for extragalactic objects, over any frequency band. These results are enabled by new data sets from the latest generation of high-resolution radio interferometers. Notably, SKA precursor and pathfinder instruments like MeerKAT and ASKAP are now able to provide us with an unprecedented level of sensitivity and resolution, which can be utilised to probe the inner regions of various astronomical objects in incredible detail. Since these regions are also where we should see the strongest DM emissions, the precise nature of the data can clearly be leveraged to provide compelling new results. 

A popular astronomical search target for indirect detection studies has been the (dwarf-spheroidal) satellite galaxies of the Milky Way, due to the large predicted emissions from their relative vicinity and highly-concentrated DM halos. However, there has also been some renewed interest in recent years for searches in galaxy clusters (GCs)~\cite{stormSynchrotronEmissionDark2017,kiewConstraintsDarkMatter2017,chanConstrainingAnnihilatingDark2020,beckGalaxyClustersHigh2023}. The motivations for GCs as search targets are that their mass is dominated by a large DM component, and their large physical size prevents annihilation products from diffusing away before they emit synchrotron radiation. The modelling of the DM halo and the diffusive environment, and thus the simulation of radio synchrotron emissions from within, is also much less uncertain than it is for smaller dwarf-spheroidal (dSph) galaxies. The main issue with these objects is their large baryonic background emissions -- usually in the form of radio halos or relics -- which makes disentangling the DM signal difficult. Thus, high-resolution and high-sensitivity observations of a GC target void of any strong diffuse radio emissions presents us with a valuable opportunity to hunt for a DM signal. 

With the above motivations, we now analyse observational data of the \RXC{} cluster, made available through the first public data release of the MeerKAT Galaxy Cluster Legacy Survey (MGCLS), completed in November 2021~\cite{knowlesMeerKATGalaxyCluster2022}. This survey includes L-band observations of 115 galaxy clusters, with full visibilities and uncertainties at $\sim$ 8`` resolution and $\sim 3-5 \;\mu\textrm{Jy/beam}$ sensitivities (for full details, see~\cite{knowlesMeerKATGalaxyCluster2022}). \RXC{} is located at z = 0.060, and lacks any significant diffuse components in the MeerKAT image maps, which are taken at a frequency of 1280 MHz. We proceed by calculating the model radio emissions from a DM component in this source, predicted through a set of reasonable parameters, and create a set of upper limits on the possible annihilation cross-section based on the comparison of this model to the observed data. 

The structure of this paper is as follows: in Section~\ref{sec:model} we outline our calculations of the DM halo structure and diffusive environment, and in Section~\ref{sec:data} we describe the observational data and the statistical approach to calculating the results. We then present the main result of this work in Section~\ref{sec:results}, which is followed by a brief discussion and concluding remarks in Section~\ref{sec:discussion}.

\section{Emission Modelling}\label{sec:model}

\subsection{Halo environment}

We model the DM halo in \RXC{} using the Hernquist-Zhao density profile~\cite{navarroStructureColdDark1996}, using mass and concentration parameters found from~\cite{shakouriATCAREXCESSDiffuse2016} and \cite{foexSubstructuresGalaxyClusters2019}, respectively. The form of this profile can be written as 
\begin{equation}\label{eqn:halo}
	\rho(r) = \frac{\rho_s}{\left(\frac{r}{r_s}\right)^{\alpha_z}\left(1+\frac{r}{r_s}\right)^{3-\alpha_z}} \; ,
\end{equation}
where $r$ is the radius from the centre of the halo and $r_s$, $\rho_s$ are scale parameters which are calculated from the above-mentioned sources. For comparative purposes, we take the halo index $\alpha_z$ to equal two values; $\alpha_z = 1$ which describes an NFW (cuspy) profile and $\alpha_z = 0.5$ which describes a more shallowly cusped profile. We then consider the constituent DM particles in the halo to be in a class of generic Weakly-Interacting Massive Particles (WIMPs), which are expected to annihilate and produce a population of kinematically-accessible Standard Model (SM) products. In this work, as we are interested in the production of electrons and positrons (hereafter referred to as just electrons), the annihilation process can be represented symbolically as $\chi\chi \rightarrow S \rightarrow e^-e^+$, where WIMPs are represented by $\chi$ and intermediate channels by $S$. As in most indirect detection studies, we consider intermediate channels individually. The distribution of electrons produced by a single annihilation, commonly referred to as the source term ($Q$), can then be calculated by
\begin{equation}\label{eqn:source}
	Q(E,r) =  \cs \diff{N_{\mathrm{e}}}{E} \mathcal{N}_{\chi}(r) \,,
\end{equation}
where $\cs$ is the usual velocity-averaged annihilation cross-section, $\inlinediff{N_e}{E}$ is the particle energy spectrum (here obtained from~\cite{cirelliPPPCDMID2011}), and $\mathcal{N}_{\chi}(r)$ is the WIMP pair density within the halo. This quantity is simply found using the DM halo density, as $\mathcal{N}_{\chi} = \frac{1}{2}\left(\frac{\rho}{M_{\chi}}\right)^2$, where the prefactor of 1/2 corresponds to the case of Majorana-like WIMPs. 

The source term $Q$ is now used to describe the continual injection of electrons into the halo, and we then model the evolution of this population using a standard cosmic ray transport equation. The dominant effects in this scenario are spatial diffusion and energy-losses, which can be encapsulated by the following equation
\begin{multline}\label{eqn:propagation}
    \pdiff{\psi(\mathbf{x},E)}{t} = \nabla\cdot\left(D(\mathbf{x},E)\nabla\psi(\mathbf{x},E)\right)\\
	+\pdiff{}{E}\left(b(\mathbf{x},E)\psi(\mathbf{x},E)\right) + Q(\mathbf{x},E) .
\end{multline}
Here $\psi$ represents the equilibrium electron distribution in the halo and $D$, $b$ represent the diffusion and energy-loss effects. These functions generally depend on the energy of the electrons $E$ as well as their position inside the halo $\mathbf{x}$ (which reduces to $r$ in the case of spherical symmetry). The diffusion term is given by 
\begin{equation}\label{eqn:diffusion}
	D(r,E) = D_0\left( \dfrac{d_0}{1\, \mathrm{kpc}} \right)^{\frac{2}{3}}\left( \dfrac{B(r)}{1\, \mu\mathrm{ G}} \right)^{-\frac{1}{3}}\left( \dfrac{E}{1\, \mathrm{GeV}} \right)^{\frac{1}{3}} ,
\end{equation}
\noindent and the energy-loss term by
\begin{align}\label{eqn:loss}
	b(r,E) &= b_{\mathrm{IC}}\left(\dfrac{E}{1\,\mathrm{GeV}}\right)^2 + b_{\mathrm{sync}}\left(\dfrac{E}{1\,\mathrm{GeV}}\right)^2B(r)^2 \nonumber \\
	& + b_{\mathrm{coul}}n_\mathrm{e}(r)\left(1+\dfrac{1}{75}\log\left(\dfrac{\gamma}{n_\mathrm{e}(r)}\right)\right) \nonumber \\
	& + b_{\mathrm{brem}}n_\mathrm{e}(r)\left(\dfrac{E}{1\,\mathrm{GeV}} \right) .
\end{align}
The parameters for these functions are defined as follows. In Eqn.~\ref{eqn:diffusion}, we use a diffusion coefficient of $D_0 = 3\times10^{28} \,\mathrm{cm^2\,s^{-1}}$. Although there is some uncertainty in this choice for extragalactic targets, we note that the final results are somewhat robust to variations in this value, differing by less than 5 per cent for changes in $D_0$ of an order of magnitude. This is likely due to the physical size of the target, as we see with galaxy clusters in general~\cite{colafrancescoMultifrequencyAnalysisNeutralino2006}. We then select the coherence length of the magnetic field $d_0 = 2.0$ kpc, where $B(r)$ is the magnetic field strength at $r$ (defined below). 

In Eqn.~\ref{eqn:loss}, each term describes a different energy loss mechanism, labelled by the subscripts on the $b$ coefficients. These are, with their corresponding values (in units of $10^{-16} \,\mathrm{GeV\,s^{-1}}$): Inverse Compton Scattering from CMB photons ($0.25(1+z)^4$), synchrotron emissions (0.0254), Coulomb scattering (6.13) and bremsstrahlung (4.7). The factor of $\gamma = E/m_ec^2$, where $m_e$ is the electron mass.

The remaining quantities, the gas density $n_e(r)$ and magnetic field $B(r)$, are modelled here with radial beta profiles, motivated by the X-ray data of \RXC{} in~\cite{crostonGalaxyclusterGasdensityDistributions2008}. This profile can be written generally as 
\begin{equation}\label{eqn:gas}
	X(r) = n_0\left[1+\left(\dfrac{r}{r_s}\right)^2\right]^{3\beta/2} \,
\end{equation}
where $X(r)$ could be either $n_e(r)$ or $B(r)$, and the normalisation and scaling parameters $n_0, r_s$ and $\beta$ are found for each profile individually. After a least-squares fit to the data in~\cite{crostonGalaxyclusterGasdensityDistributions2008}, we adopt the set of parameters laid out in Tab.~\ref{tab:params}.
\begin{table}[h]
	\centering
	\tbl{List of parameters for the radial beta profiles of $n_e(r)$ and $B(r)$.\label{tab:params}}{%
	\begin{tabular}{ r | c  c }
		 & \multicolumn{2}{ c }{Values} \\[0.25em]
		Parameters & $n_e(r)$ & $B(r)$  \\
		\toprule
		$n_0 $ & $3.95\times10^{-3} \, \mathrm{cm^{-3}}$ & $5.00 \, \mu\mathrm{G}$ \\
		% $n_{0,2}$ & $1.19\times10^{-3} \, \mathrm{cm^{-3}}$ & $5.00 \, \mu\mathrm{G}$ \\
		$r_s$ (Mpc) & $0.11$ & $0.11$ \\
		% $r_{s,2}$ (Mpc) & $0.25$ & $0.25$ \\
		$\beta $  & $-0.81$ & $-0.40$ \\
		% $\beta_{s,2}$ & $-0.87$ & $-0.43$ \\[0.5em]
	\end{tabular}}
\end{table}

The form of the transport equation (Eqn.~\ref{eqn:propagation}) is now fully specified, and we solve it numerically using the method laid out in~\cite{beckGalaxyClustersHigh2023} (therein referred to as the \textit{ADI} method). This involves the discretisation of Eqn.~\ref{eqn:propagation}, and an iterative solution method that makes use of a generalised Crank-Nicolson scheme in each dimension ($r$ and $E$).  For details regarding this technique, we refer the reader to~\cite{beckGalaxyClustersHigh2023}, as well as~\cite{strongPropagationCosmicRay1998,regisLocalGroupDSph2015,evoliCosmicrayPropagationDRAGON22017}, wherein similar methods have been successfully implemented to solve the transport equation.

\subsection{Synchrotron emissions}

After finding the equilibrium distribution of electrons, we calculate the synchrotron radio emissivity by
\begin{equation}\label{eqn:emm}
j_{\mathrm{sync}} (\nu,r) = \int_{0}^{M_\chi} \d E \, \psi_{e^{\pm}}(E,r) P_{\mathrm{sync}} (\nu,E,r) \; ,
\end{equation}
where $\nu$ is the synchrotron frequency, $\psi_{e^{\pm}}$ is the sum of electron and positron equilibrium distributions and $P_{\mathrm{sync}}$ is the emitted synchrotron power of an electron with an energy of $E$ (this is calculated as in~\cite{beckRadioFrequencySearchesDark2019}). With this, we finally calculate the surface brightness 
\begin{equation}\label{eqn:sb}
I_{\mathrm{sync}} (\nu,R) = \int \d l \, \frac{j_{\mathrm{sync}}(\nu,\sqrt{R^2+l^2})}{4 \pi} \; , 
\end{equation}
where $l$ is the line-of-sight to a point in the halo at radius $R$. The values calculated here are then mapped onto a FITS image, in order to be used in the statistical analysis of the data (described in Sec.~\ref{sec:data}).

\section{Observational data}\label{sec:data}

\begin{figure}[h]
	\centering
    \includegraphics[width=0.48\textwidth]{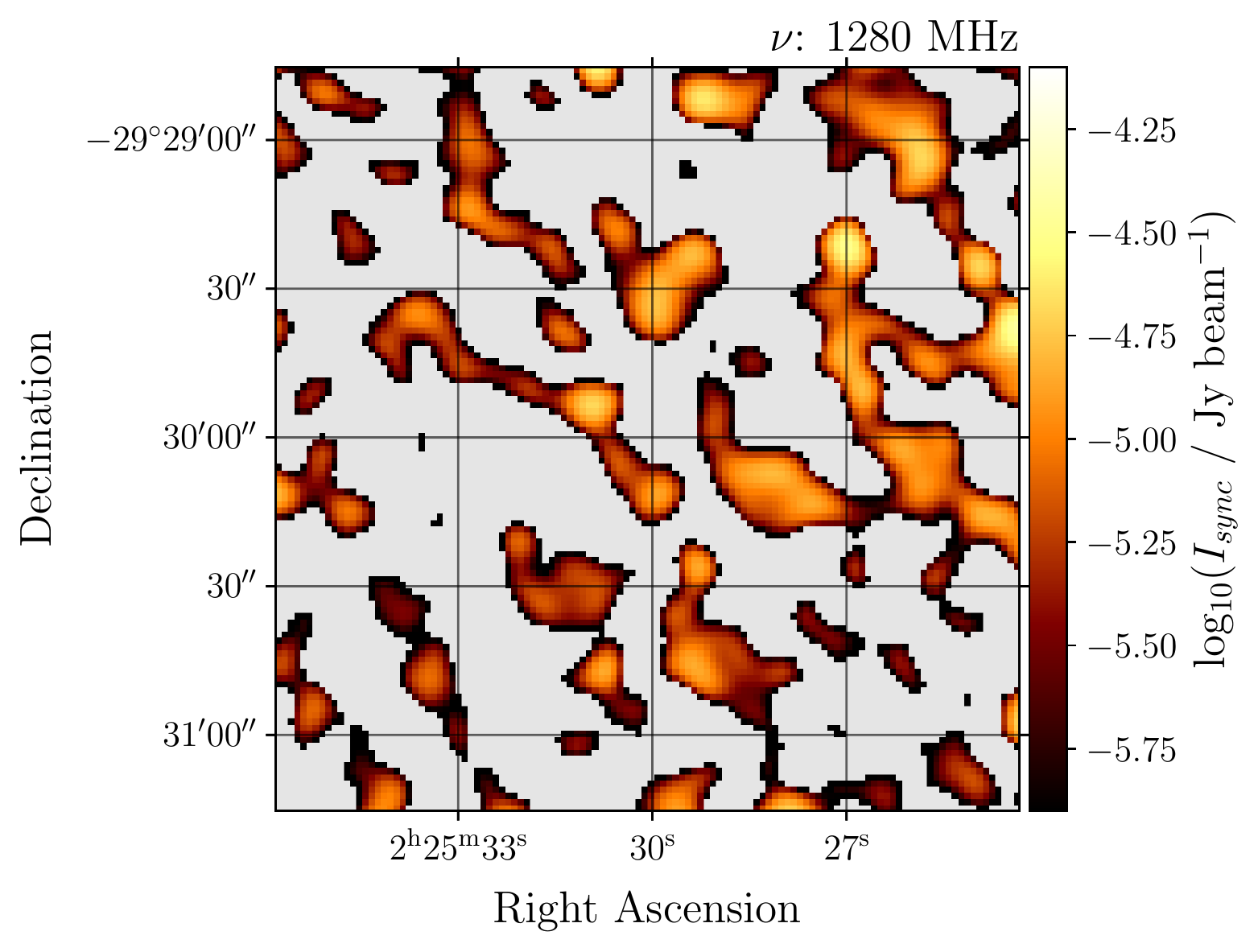}
	\caption{A representation of the \RXC{} surface brightness data obtained from~\cite{knowlesMeerKATGalaxyCluster2022}, after source-subtraction and with negative pixels masked (shown in light gray.)}\label{fig:data}
\end{figure}

The full set of image data for \RXC{} was obtained from the MGCLS data-release website \footnote{http://mgcls.sarao.ac.za/data-releases/}, whereafter point source emissions were subtracted from the surface brightness data with the use of the \texttt{PyBDSF}\footnote{https://pybdsf.readthedocs.io/en/latest/} code package, using default input parameters. We further restricted the image to a 2.5' x 2.5' square region around the cluster centre, at the MeerKAT pointing coordinates of $\mathrm{(RA,Dec)_{J2000}} = (36.3750\degree,-29.500\degree)$, and then masked any remaining negative pixels. This region, shown in Fig.~\ref{fig:data}, contains $N \sim 5\times 10^3$ good pixels, and with a beamwidth of $\sim 8.1''$ x $\sim 7.8''$ this corresponds to a usable area of roughly a hundred beamwidths.  

With our DM emission model projected onto the same sky coordinates as the data, we then proceed to perform a standard likelihood-ratio test with $\cs$ as a free parameter. Given the large set of usable pixels in the image, we assume a Gaussian form for the likelihood $\mathcal{L}_i = \mathrm{e}^{-\chi^2}/2$, where 
\begin{equation}\label{eqn:chi2}
	\chi^2 = \dfrac{1}{N_{\mathrm{b}}}\sum_{i=1}^N \left(\dfrac{E_i-O_i}{\sigma_i}\right)^2 \; 
\end{equation}
is the $\chi^2$ statistic, and $O_i$, $E_i$ and $\sigma_i$ represent the corresponding data and model emissions with their related uncertainties ($\sim~5\,\mu \mathrm{Jy}$ per pixel). We follow~\cite{regisLocalGroupDSph2015} by weighting the statistic by the number of pixels per beam ($N_{\mathrm{b}}$), to account (approximately) for the correlation between pixels. Here we are assuming that the pixels within the FWHM of a beam are correlated, while those outside are not.

We then perform the likelihood ratio test with the result of Eqn.~\ref{eqn:chi2}, where $2\ln\left(\mathcal{L}_i / \mathcal{L}_{\mathrm{max}} \right) = \chi^2_i - \chi^2_{\mathrm{min}} \equiv  \chi_c$. Here $\chi_c$ represents the one-sided confidence level in the cumulative distribution function
\begin{equation}\label{eqn:prob}
	P = \int^{\infty}_{\chi_c}\; d\chi \frac{\exp(-\chi^2/2)}{\sqrt{2\pi}} \;.
\end{equation}
In this work we find $2\sigma$ exclusion values, which corresponds to $P = 1-\alpha$, where $\alpha = 0.95$. We select the values of $\cs$ that satisfy this condition for a set of WIMP masses, which are presented in the following section. 

\section{Results}\label{sec:results}
\begin{figure}[h]
	\centering
    \includegraphics[width=0.48\textwidth]{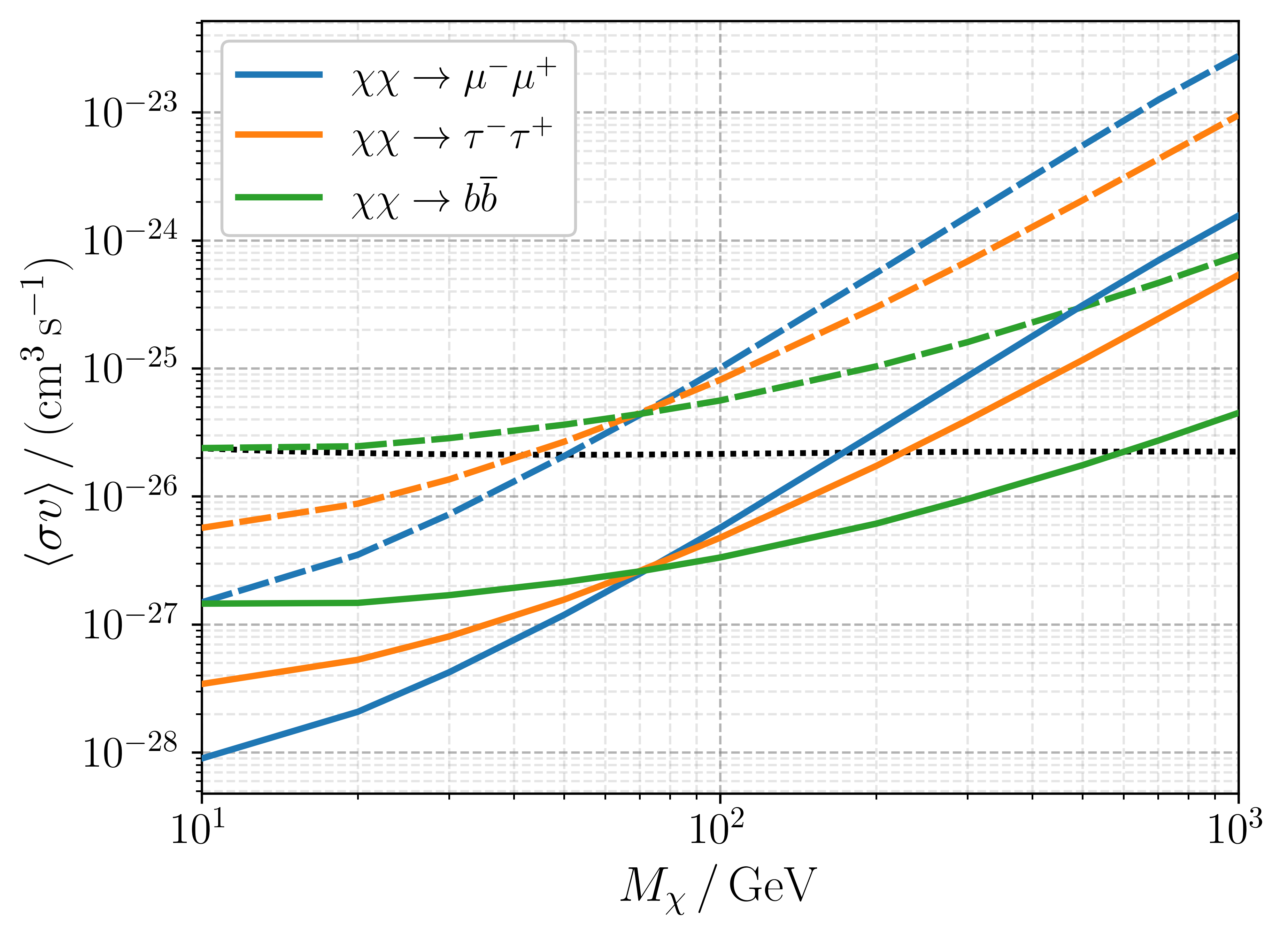}
	\caption{Upper limits ($2\sigma$) on the annihilation cross-section for WIMPs in \RXC{} annihilating via three intermediate channels. The solid and dashed lines represent results with $\alpha_z=1$ and $\alpha_z=0.5$ respectively, and the dotted line shows the calculated thermal relic value for the cross-section, taken from~\cite{steigmanPreciseRelicWIMP2012}.}\label{fig:limits}
\end{figure}

The main result of this work -- the outcome of the exclusion analysis based on the likelihood ratio test described in Sec.~\ref{sec:data} -- is presented in Fig.~\ref{fig:limits}. The curves represent the intermediate annihilation channels of bottom quarks ($b\overline{b}$), muons ($\mu^-\mu^+$) and tau leptons ($\tau^-\tau^+$), and are representative of the larger set of total possible channels. 

In Fig.~\ref{fig:limits} we have displayed the results for the two halo density profiles mentioned in Sec.~\ref{sec:model}. There is a significant difference between each (a factor of $\sim~16$), which is a consequence of the relative cuspiness of each profile. The more cuspy profile of $\alpha_z = 1$ contains higher densities in regions $r \lesssim r_s$, which leads to stronger emissions therein and thus stronger contraints on the value of $\cs$. While the exact nature of the density profile in this source is not currently known, there is evidence that favours NFW-like profiles in galaxy clusters (see for example the discussion in~\cite{beckGalaxyClustersHigh2023}). We thus consider the shallowly cusped profile of $\alpha_z=0.5$ as an upper bound for the uncertainty in the nature of the halo profile.

\section{Discussion and conclusions}\label{sec:discussion}

The constraints on the DM annihilation cross-section found in this analysis using a cuspy NFW halo profile are comparable to those found in~\cite{regisEMUViewLarge2021}, being within the uncertainty bounds for each channel. When considering a shallowly-cusped halo profile we find constraints that are significantly weaker, however still comparable to those found in the literature for various other targets (for example in~\cite{egorovUpdatedConstraintsWIMP2022,kiewConstraintsDarkMatter2017,regisLocalGroupDSph2014,chanRulingOut1002019}). We note that the use of GC targets in these analyses has a significant advantage over more common targets like galaxies or dSphs -- the uncertainty in the modelling of physical parameters is greatly reduced. When combined with high-resolution observations of a cluster that lacks a notable diffuse component (such as \RXC{}), the resulting limits are more robust to these uncertainties.

The statistical analysis of the radio data adopted here ( described in Sec.~\ref{sec:data}) is similar to the technique used in~\cite{regisEMUViewLarge2021}. This approach, that uses the morphology of the source (through a pixel-by-pixel analysis of the data), together with the high sensitivity of the ASKAP images, is argued to account for the large improvement in the constraints compared to previous LMC observations. In the case of \RXC{} MeerKAT data we note that while the number of usable data points in the image is still relatively large at $\sim 5 \times 10^3$, there is a significant proportion of the total number of pixels in the image that have negative flux values. This is mainly due to systematic artefacts, however there is a small fraction coming from over-subtraction during the removal of point sources. In this case the inner regions of the image ($\lesssim 3$ beam radii) are free from these over-subtracted negative pixels, while outer regions contain progressively more. These pixels are unusable in this analysis and are thus masked (i.e. they do not enter into the sum of Eq.~\ref{eqn:chi2}). Future studies of similar datasets might thus benefit from a point-source subtraction procedure that is tailored for the morphology of the source image, together with a more optimal or quantitative criteria for the selection of the region of interest in the image. Further, we note that tailored observations of a source (i.e. those not from a set of survey results) might have lower associated uncertainties from longer exposure times. This would have a direct improvement on the strength of the upper limits for $\cs$. 

With the availability of high-resolution data from current radio interferometers, and in anticipation for next-generation instruments like the SKA, we expect the constraining power of radio DM searches to improve greatly. To realise this potential, rigorous statistical analyses need to be coupled with accurate modelling of the diffusive and halo environment. In the case of the \RXC{} data presented here, we are able to isolate the inner 2.5' x 2.5' region of the image and still utilise a significant number of data points in the analysis. The use of a small central region like this allows us to ignore the outer regions of the DM halo that produce weaker emissions, and thus contribute less to the exclusion limits. The modelling of this region is thus of vital importance. Here we make use of a numerical solution to the electron transport equation which allows us to incorporate the spatial dependance of the magnetic field and thermal gas density into the solution. This should further reduce the level of uncertainty our results have due to physical parameters in the modelling process.

In conclusion, we have calculated a set of $2\sigma$ exclusion limits for the annihilation cross-section of generic WIMPs, using the radio observations of the \RXC{} galaxy cluster from the MGCLS. With our most constraining scenario of a cuspy NFW halo profile, the results are competitive with the most stringent limits yet found for generic WIMPs. Based on these results, we look forward to the upcoming datasets produced by radio interferometry instruments and their corresponding science teams, which promise to be a vital aid in the ongoing search for DM.

\section*{Acknowledgements}

This work is based on the research that was supported by the National Research Foundation of South Africa (Bursary No. 112332). G.B. acknowledges support from a National Research Foundation of South Africa Thuthuka grant no. 117969. NL acknowledges the financial assistance of the South African Radio Astronomy Observatory (SARAO) towards this research (www.sarao.ac.za). \\

\noindent The authors declare no conflict of interest. 

\bibliographystyle{unsrt}
\bibliography{references/NuDM2022}

\end{document}